\documentclass[journal]{IEEEtran}
\usepackage{xcolor,soul,framed}

\colorlet{shadecolor}{yellow}
\usepackage[pdftex]{graphicx}
\graphicspath{{../pdf/}{../jpeg/}}
\DeclareGraphicsExtensions{.pdf,.jpeg,.png}
\usepackage{graphicx}
\usepackage{caption}
\usepackage[export]{adjustbox}
\usepackage{wrapfig}
\usepackage[skip=2pt,font=footnotesize]{caption}
\usepackage{hyperref}
\hypersetup{
    colorlinks=true,
    linkcolor=blue,
    filecolor=blue,      
    urlcolor=black,
    citecolor =blue,
    pdftitle={Design of Perfect anomalous reflector},
    pdfpagemode=FullScreen
    }

\usepackage{subfig}
\usepackage[cmex10]{amsmath}
\usepackage{array}
\usepackage{mdwmath}
\usepackage{mdwtab}
\usepackage{eqparbox}
\usepackage{url}
\usepackage{textcomp}
\usepackage{gensymb}
\usepackage{multirow}
\usepackage{multicol}
\usepackage{stfloats}
\usepackage{mathrsfs}
\usepackage[noadjust]{cite}
\usepackage[ragged]{sidecap}
\usepackage{bm}

\begin{document}

\title{Simultaneous Angle-of-Arrival Sensing and Anomalous Deflection with Aperiodically Loaded Patch Arrays}

\author{
Mostafa~Movahediqomi,
Sravan~K.~R.~Vuyyuru,~\IEEEmembership{Member,~IEEE,}
Grigorii~Ptitcyn,
Risto~Valkonen,~\IEEEmembership{Member,~IEEE,}\\
Viktar~S.~Asadchy,~\IEEEmembership{Senior Member,~IEEE,}
Do-Hoon~Kwon,~\IEEEmembership{Senior Member,~IEEE,}
and Sergei~A.~Tretyakov,~\IEEEmembership{Fellow,~IEEE}

\thanks{This work was supported in part by the European Union’s Horizon 2020 MSCA-ITN-METAWIRELESS project (the Marie Skłodowska-Curie grant agreement No 956256),  the US Army Research Office under grant W911NF-19-2-0244, the Research Council of Finland and DoD (Project No. 365679), and the Finnish Foundation for Technology Promotion.}
\thanks{M.~Movahediqomi, V.~S.~Asadchy, and S.~A.~Tretyakov are with the Department of Electronics and Nanoengineering, School of Electrical Engineering, Aalto University, 02150 Espoo, Finland (e-mail: mostafa.movahediqomi@aalto.fi, viktar.asadchy@aalto.fi, sergei.tretyakov@aalto.fi).}
\thanks{S.~K.~R. Vuyyuru is with the Department of Electronics and Nanoengineering, School of Electrical Engineering, Aalto University, 02150 Espoo, Finland, and was with Nokia Bell Labs, Karakaari 7, 02610 Espoo, Finland (e-mail: sravan.vuyyuru@aalto.fi).}
\thanks{G.~Ptitcyn is with the Department of Electrical and Systems Engineering, University of Pennsylvania, Philadelphia, PA 19104, U.S.A. (e-mail: ptitcyn@seas.upenn.edu).}
\thanks{R.~Valkonen is with Nokia Bell Labs, Karakaari 7, 02610 Espoo, Finland (e-mail: risto.valkonen@nokia-bell-labs.com).}
\thanks{D.-H.~Kwon is with the Department of Electrical and Computer Engineering, University of Massachusetts Amherst, Amherst, MA 01003, USA (e-mail: dhkwon@umass.edu).}
}  



\maketitle
\begin{abstract}
 We propose and numerically demonstrate a reconfigurable patch antenna array that enables simultaneous incident wave sensing and anomalous reflection without prior knowledge of the propagation environment. We acquire anomalous reflection by suppressing parasitic scattering through accurate and efficient optimization of induced load currents and by varying impedances of reactive loads. By mitigating parasitic scattering lobes, we demonstrate the feasibility of accurately detecting the incoming illumination angle via the spatial Fourier transform of the optimized load current distribution, facilitated by tunable reactive loads. This approach eliminates the need for additional RF chains, pre-computed data, or calibration measurements. The developed strategy, which integrates arithmetic load optimization with angle-of-arrival sensing, is applicable to general finite-size arrays.

\end{abstract}


\IEEEpeerreviewmaketitle


\section{Introduction}\label{sec:Intro}

\IEEEPARstart{I}n recent years, reconfigurable intelligent surfaces (RISs), as two-dimensional structures, have been designated to engineer the electromagnetic (EM) environment for the next generation of mobile networks (6G). Through this new platform, the goal is to dynamically alter the reflection wavefront characteristics of reflective surfaces, allowing real-time adaptation to variations of the propagation environment and network layout. The feasibility of this approach is guaranteed through the tuning of electromagnetic features of reflective surfaces. One straightforward way is to reconfigure the load impedances at the ports of the array elements that comprise an RIS~\cite{Smart_Radio_Environments,MarcoCommModelsRIS,IntMetaLiu2019,movahediqomi2023comparison,vuyyuru,li2023tunable,vuyyuru2024finite,renzo2019smart,di2020reconfigurable}.

In dynamic communication networks, the incoming wave angle from the mobile user to RIS is unknown, making it one of the prominent tasks for the future-generation network to sense that unknown angle. In other words, the Angle-of-Arrival (AoA) needs to be sensed while simultaneously reflecting the waves into the direction of the receiver (towards the downlink receiver, i.e., the base station (BS)). These two functionalities, referred to as integrated sensing and communication (ISAC), can be achieved using a single electromagnetic (EM) scattering surface. Numerous studies have been conducted to implement ISAC through various methodologies. Among the first studies, antennas were used as tools for AoA estimation using Fourier techniques, but the work was focused primarily on channel analysis~\cite{kalliola2000real}. Recently, ISAC has been realized by using RISs with limited functionalities. In Refs.~\cite{wang2020channel,zheng2021efficient} the authors developed an RIS-based structure to send the communication pilots between the BS and user equipment (UE), but the main drawback of the proposed approach is its inability to decouple BS and UE, which needs to be mitigated by incorporating an additional RF chain and antenna dedicated to the sensing part. Introducing additional components into a system will diminish the effective coverage area, and the extra RF chain increases the structural complexity and the overall cost. A notable proposal emerged in Ref.~\cite{liaskos2019absense} that suggests absorption of incoming EM waves using an RIS equipped with a sensor system, which measures the absorbed energy against a pre-established lookup table. However, its practical applicability is constrained by the low resolution of existing datasets. A novel sensing paradigm has recently emerged, utilizing a portion of the total wave power for sensing by directing it through a waveguide positioned beneath the RIS~\cite{alexandropoulos2023hybrid,zhang2023channel,zhang2021channel,alamzadeh2023detecting,alamzadeh2021reconfigurable,alamzadeh2022sensing,albanese2022marisa,albanese2023ares,jiang2023simultaneously}. However, the waveguide sensing method has inherent limitations. The sensing process occurs over the entire RIS interface domain, including incident and scattered EM fields, raising the challenge of separating these fields and complicating the detection of the direction of the incident waves, thereby necessitating multiple pre-computed test RIS illuminations. Moreover, this technique is constrained to a single plane-wave illumination because of its waveguide's geometry. Additionally, post-sampling RF combiners are necessary to aggregate the signals before their transmission to RF detectors, which adds complexity and raises cost issues. Another study merged the sensing of the AoA with radar cross-section (RCS) manipulation using space-time modulated metasurfaces \cite{fang2023multifunctional}. The estimation was conducted in the transmission region (behind the metasurface), while the RCS manipulation occurred in the reflection region (in front of the metasurface). However, signal processing is needed to do the AoA estimation behind the metasurface when the transmitted harmonics reach the receiver antenna used for sensing. Multiple pre-computed RIS illuminations are often required, typically leveraging deep neural networks~\cite{lin2021single,huang2022machine,chen2022artificial}. However, this approach is inherently constrained to single plane-wave illumination. Extending it to support simultaneous multiple-wave excitations would necessitate impractically large datasets, as the data size grows exponentially with the number of incident waves. Additionally, other works have implemented ISAC using periodic metasurfaces and structures with specific electromagnetic properties \cite{lin2021single,huang2022machine,chen2022artificial}. Paper~\cite{huang2024comprehensive} provides an in-depth review of metasurface-assisted AoA estimation techniques. 

\begin{figure*}[t]
\centering
\includegraphics[width=7in,height = 2.8in]{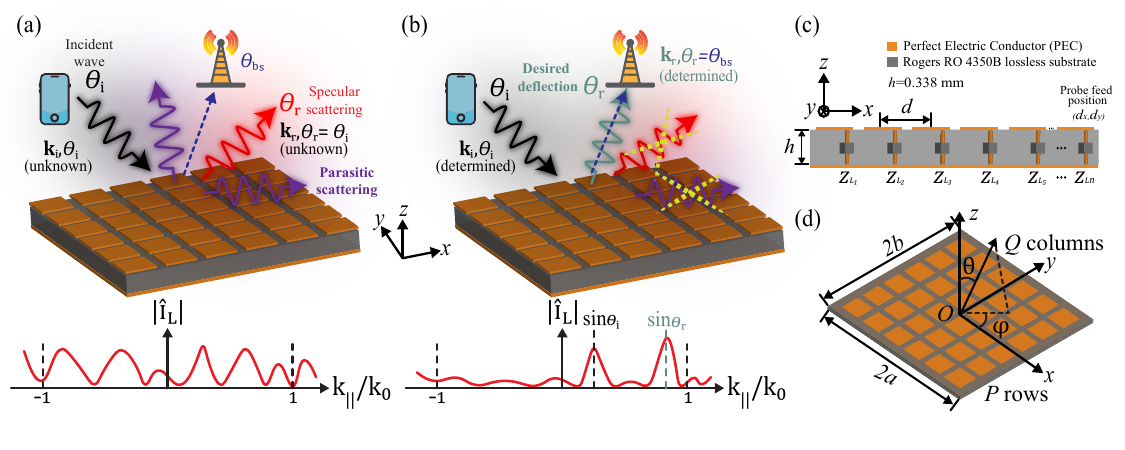}
\caption{The 3-D schematic illustration depicts the proposed dual-functional planar RIS comprising loaded patch elements on a finite grounded dielectric substrate of a thickness $h$, under a plane-wave illumination at unknown angle $\theta_{\rm{i}}$ and the desired deflection angle at known $\theta_{\rm{r}}$. The lowermost plot depicts the Fourier transform of surface averaged electric current, $|\widehat{\mathbf{I}}_{\rm {L}}|$, passing through the load terminals along a specific direction. The stages depict (a) pre-optimization, characterized by specular and parasitic reflections from an unknown angle due to the random distribution of load reactances, and (b) post-optimization, achieving both desired anomalous reflection and AoA estimation. The bottom plot, picturing the spatial Fourier transform of the optimized current, highlights a dominant peak in the desired direction, while the second peak reveals the AoA of the incident wave. (c) A side view of the RIS indicating the equally spaced patch elements with a period $d$. (d) The coordinate system represents the RIS with $p$ rows and $q$ columns, and the array spans around the coordinate system origin $O$.}
\label{fig1:conceptual_theory}
\end{figure*}

Conversely, in~\cite{Mostafa24SensingWires}, an alternative approach was proposed for achieving AoA sensing under arbitrary illuminations by leveraging the measured data already used to optimize power reflection toward the receiver. The underlying concept behind this approach is a straightforward calculation of the Fourier transform of the  load currents followed by an optimization of a proper spatial harmonic that corresponds to  reflection in the desired direction. Once optimization is performed, the spatial Fourier spectrum reveals two notable peaks. The first peak corresponds to the tangential wavenumber associated with the desired reflection angle, while the second peak within the propagation range corresponds to effectively suppressed specular reflections from the ground plane. Subsequently, this wavenumber can be associated with the direction of the specular reflection, revealing the AoA. If the RIS is illuminated not by a single plane wave, the spatial spectrum of the load currents reveals the angular spectrum of the incident beam. 
In practical implementation, by employing amplitude detectors connected to each load and phase detectors between each pair, the current at each terminal can be measured as a single complex number, allowing for real-time tracking. The main difference between that work and the previous studies is the nature of the proposed structure in terms of EM properties. In ~\cite{Mostafa24SensingWires} the geometry of the structure is periodic, while the current distribution is not. This gives access to arbitrary spatial harmonics, therefore enabling reflection (reception) into (from) arbitrary directions, unlike structures with periodical electromagnetic properties that are not capable of this functionality. In other words, in~\cite{Mostafa24SensingWires}, the reflectarray antenna itself is used as a network sensor and estimation of AoA is performed alongside with the optimization of reflection. The angular resolution is determined by the RIS dimensions, wherein a larger aperture results in a narrower beamwidth and improved resolution, as  defined by the classical antenna theory. 

The concept of~\cite{Mostafa24SensingWires} was theoretically demonstrated in a model finite arrays of infinite impedance-loaded wires, for which analytical formulation of the fields is well known. However, arrays of wires are not a practical platform for actual applications due to the hard implementation process of very thin wires floating above a ground plane (without a dielectric substrate). However, therefore in this work, we extend this concept to patch arrays, i.e., geometrically periodic reflectors composed of reactively loaded patch elements within a supercell.  An arithmetic optimization method developed in ~\cite{vuyyuru,vuyyuru2024modeling} can be  utilized to design theoretically infinite periodic arrays capable of efficient scanning, including extreme angles. However, periodic structures scan only at discrete angles corresponding to the supported Floquet harmonics with fixed periodicity for specific incident and reflection angles, ignoring edge element effects. In~\cite{vuyyuru23array,vuyyuru2024finite}, instead of supercell-level optimization, an algebraic-level global optimization approach for a finite array is utilized to achieve continuous wide-angle scanning with high efficiency. These optimization methods achieve an optimal power reflection efficiency even for deflecting reflected waves at extreme angles, contrasting with the poor efficiency of the conventional phase gradient approach and lacking the AoA detection capability in patch arrays.

This study builds upon the theoretical concept introduced in Ref.~\cite{Mostafa24SensingWires}, focusing on developing an ISAC application using an aperiodic-loaded array of square patches on a finite-size grounded substrate. Specifically, our primary aim is to maximize the power reflected to the desired direction by utilizing different optimization objective functions together with determining the unknown angle or angles of arrival. A key challenge in contrast to the work presented in in~\cite{Mostafa24SensingWires} is the inability to rely solely on analytical solutions. Instead, we incorporate simulated data for the array parameters (the impedance matrix) combined with transmitting-receiving antenna theory to compute the scattered field. Due to the aperiodic nature of the array, this technique allows simultaneous redirection of incoming waves to arbitrary angles with high power efficiency and AoA sensing.
 
This paper is organized as follows. Section~\ref{sec:methodology} introduces two distinct optimization techniques based on the scattering synthesis principle for finite patch antenna arrays with a subwavelength geometrical period and the AoA detection methodology. Section~\ref{sec:results} focuses on optimized finite arrays with different element spacings with aperiodic loadings to realize simultaneous anomalous deflection and AoA detection. A general summary and concluding observations are drawn in Section~\ref{sec:conclusion}.

\section{Concept and Formulation}\label{sec:methodology}

\subsection{Governing Idea}\label{subsec:core_idea}

Here, we consider an uplink communication network comprising two links (i.e., two-path), as depicted in Fig.~\ref{fig1:conceptual_theory}. The first link connects the UE to the RIS, with the incoming wave arriving from an unknown direction and illuminating the RIS array. Subsequently, another link is established between the RIS and the uplink receiver (i.e., BS). Here, the receiver's location is predetermined and known, and the reflected wave from the RIS partially reaches the BS. For an arbitrary array, the current distribution over the patch array can be characterized by its spatial Fourier spectrum
\begin{equation}
\label{Eq:transform}
\widehat{\bf I}_{\rm L} ({\bf k}_{||}) = \int_{-\infty}^{\infty} \mathrm {\bf I}_{\mathrm L}(\rho) e^{-j{\bf k}_{||}\cdot {\bm \rho}} \, {\rm d}\rho, 
\end{equation}
where ${\bm \rho}=x\hat x+y\hat y$ and ${\bf k}_{||}=k_x\hat x+k_y\hat y$. Here ${\bf k}_{||}$ is the tangential wavenumber and has the relation with the free-space wavenumber $k$ such that $\sin\theta= |{k}_{||}|/k$. The angle $\theta $ is the angle between the wavevector of the corresponding plane-wave component and the normal to the array plane. We can utilize the current flowing through the tunable loads as an adjustable EM property to maximize the power reflected in the known desired  direction. This method enables real-time optimization, allowing adaptive tracking of illumination changes, including variations in the arrival angle. Notably, this approach does not require any prior knowledge of the AoA. The optimized current within the spatial spectrum will exhibit two distinct peaks due to the conservation of power: the first peak signifies the formation of a focused single beam redirected toward the desired receiver. The second peak emerges around the specular reflection angle to cancel out the directly reflected wave from the ground plane, which enables us to detect and estimate AoA, as depicted in Fig.~\ref{fig1:conceptual_theory}(b). 

\subsection{Scattering Synthesis Techniques for Aperiodically Loaded Anomalous Reflectors}

Figure~\ref{fig1:conceptual_theory} illustrates a dual-functional RIS illuminated by a plane wave from an unknown direction. Importantly, the concept is general and can be applied to scenarios of incident waves with arbitrary polarizations and propagating in arbitrary incidence planes.  The structure under study represents a planar array of identical square metallic patches on a grounded dielectric substrate with a height $h$. In our models, we assume that both the patches and the ground plane are made of a perfect electric conductor (PEC). This is a typical equispaced grid in the $xy$-plane, with each patch element terminated by a reconfigurable passive lumped load. The subwavelength-spaced array elements allow efficient scanning, i.e., tuning the deflection angle for various incident angles. Assuming $e^{j\omega t}$ harmonic time dependence, without loss of generality, we consider here a finite array illuminated by a plane wave with transverse electric (TE) polarization propagating in the $xz$ incidence plane:
\begin{equation}
\label{Eq:incident_E_field}
\mathbf{E}^i = \hat{y}E^i_0e^{-jk(x\sin\theta_{\rm i}-z\cos\theta_{\rm i})}, 
\end{equation}
where $E^i_0$ is the amplitude of the incident wave and $\theta_{\rm i}$ is the angle of incidence, and $k$ is the free-space wavenumber. In the following, in addition to the receiving (Rx) regime where the patch elements are loaded by passive loads, we will also consider transmitting regime (Tx) where patch antennas are excited by external voltage sources. In Fig.~\ref{fig1:conceptual_theory}(a),  $N$ patch elements forming an Rx antenna array are initially loaded with a set of random-valued $N$ tunable reactive impedance loads $Z_{Ln}~(n=1,\ldots,N)$, leading to random scattered fields and the associated nonuniform current distribution at all individual load terminals $\mathbf{I}_L$. These currents can be expressed in a column vector form 
as~\cite{vuyyuru2024finite,vuyyuru23array} 
\begin{equation}
\label{Eq:patch_load_current}
\mathbf{I}_L =\left(\overline{\overline{Z}}_\mathrm{A}+\overline{\overline{Z}}_\mathrm{L}\right)^{-1}
\mathbf{V}_\text{oc},
\end{equation}
where $\mathbf{V}_\text{oc}$ is the $N\times 1$ column vector representing voltages at the patch terminals with open-circuit loads. The load impedance matrix $\overline{\overline{Z}}_\mathrm{L}$ is diagonal with reactive load values $Z_{\mathrm{L}n}$, and $\overline{\overline{Z}}_\mathrm{A}$ is the $N\times N$ input impedance matrix that includes both self-impedances (diagonal terms) and mutual impedances (symmetrical non-diagonal terms). The total far-field scattered $E$-field, $\mathbf{E}^s(Z_{\mathrm{L}n})$, scattered by a finite antenna array in a direction $(\theta_{\rm r},\phi_{\rm r})$ is given by the sum of zero-current scattering and port-current scattering~\cite{collinspaper,vuyyuru2024finite}: 
\begin{equation}
\label{Eq:radiated_E_field}
\mathbf{E}^s(Z_{\mathrm{L}n}) = 
\mathbf{E}^s(Z_{\mathrm{L}n}=\infty)
-\sum_{n=1}^N
I_{\mathrm{L}n}(Z_{\mathrm{L}n})
\mathbf{E}_{\mathrm{I}n}(\theta_{\rm r},\phi_{\rm r})
. \end{equation}
Here, $I_{\mathrm{L}n}(Z_{\mathrm{L}n})$ and $\mathbf{E}_{\mathrm{I}n}(\theta_{\rm r},\phi_{\rm r})$ are the load port currents dependent on the load impedance $Z_{\mathrm{L}n}$ and the radiated $E$-field in the Tx mode for a unit input current excitation for predefined $(\theta_{\rm r},\phi_{\rm r})$ associated with the $n$-th port, respectively. The scattered field of the patch array with open-circuit loads is denoted as $\mathbf{E}^s(Z_{\mathrm{L}n}=\infty)$. The parameters $\overline{\overline{Z}}_\mathrm{A}$ and $\mathbf{E}_{\mathrm{I}n}(\theta_{\rm r},\phi_{\rm r})$ are computed using preliminary Tx full-wave EM simulations. Similarly, the parameter $\mathbf{E}^s(Z_{\mathrm{L}n}=\infty)$ is acquired from preliminary Rx full-wave EM simulations. Equation~\eqref{Eq:radiated_E_field} is expanded in terms of the bistatic scattering cross-section (SCS) $\sigma$ for the loaded patch antenna array~\cite{balanis2015antenna}
\begin{equation}\label{SCS}
\mathbf{\sigma}(\mathbf{Z}_L,
\theta_{\rm r}) = 
\lim_{r\to\infty}4\pi r^2
\frac{|\mathbf{E}^s(\mathbf{Z}_L)|^{2}}{|\mathbf{E}^i(O)|^{2}}.
\end{equation}
This technique effectively accounts for edge effects and mutual coupling between all radiating patches, regardless of the incident wave and load conditions. The key optimization parameters in the corresponding synthesis approach are the load impedances of the patches, allowing for continuous tunability of the desired scattering direction.

The surface electric current flowing on the patches passes through the tunable loads. Therefore, we can reconfigure our structure by considering either the currents on the patches or the tunable loads. However, selecting the current on the loads simplifies the process of solving the continuous Fourier transform integral. Since the currents flow in bulk loads, the current distribution to be measured in the optimization process can be represented as a sum of spatial Dirac delta functions, corresponding to the positions shown in Fig.~\ref{fig1:conceptual_theory}(d) for $P$ rows and $Q$ columns with period $d$ for an $N$-element array:
$I_L(x,y)=\sum_{q=1}^{Q}
\sum_{p=1}^{P} I_{Ln}
\delta(x-[q-1]d)\delta(y-[p-1]d)$,
where $n=(p-1)Q+q$.
Using the spatial Fourier transform (Eq.~\eqref{Eq:transform}), the spectrum of the load current distribution can be found by a trivial calculation of the integral of delta functions as
\begin{equation}
\label{Eq:spatial_spect}
\widehat{I}_L(k_x)
=\sum_{q=1}^Q\sum_{p=1}^PI_{Ln}e^{-jk_x(q-1)d},\
n=(p-1)Q+q.
\end{equation}

\subsection{Angle of Arrival Extraction}

In the optimized case, the patch current spectrum $\widehat{\mathbf{I}}_{\rm {L}}(k_x)$ will contain a peak corresponding to the wave scattered in the desired direction. Importantly, spectrum $\widehat{\mathbf{I}}_{\rm {L}}(k_x)$ will have an additional  strong peak at the spatial frequency $k_x=k\sin\theta_\text{i}$ which is generated naturally to cancel out the specular reflections from the ground plane. Essentially, it means that the secondary peak carries information about the direction of the incoming wave. As a result, we can detect the AoA by analysing the Fourier transform (FT) of the induced load current distribution, as shown in Fig.~\ref{fig1:conceptual_theory}(b). If the array is excited by several waves coming from different directions, there will be several corresponding peaks in the spatial spectrum, whose positions and complex amplitudes will reveal full information about the spatial spectrum of the incident field, i.e., complex phases and amplitudes of incident plane-wave components.

\section{Numerical Analysis of Optimized Patch Arrays for Dual Functionality}\label{sec:results}

In this section, we show and discuss the performance characteristics of the proposed ISAC structure. Figure~\ref{fig1:conceptual_theory}(a,b) shows a linear one-dimensional finite array made of rectangular metal patches above a grounded dielectric substrate. 
Two arrays with subwavelength element spacings ($\lambda/2$ and $\lambda/4$) are designed and compared, to allow studying the impact of the element spacing on the phenomena of anomalous reflection and AoA detection.

In the first step, a unit cell of the structure is simulated with periodic boundary conditions to determine the dimensions of a self-resonant patch antenna. This step is not mandatory for a successful design, however, it speeds up the convergence. We conduct full-wave EM numerical simulations using CST Microwave Studio. Initially, a linearly polarized PEC square patch is tuned for a $50\ \Omega$ impedance match at $28$~GHz. We employ a lossless Rogers RO4350B substrate (a relative permittivity of $\epsilon_r=3.66$ and a thickness of 0.338~mm). The optimized patch length and the load probe position $(d_x,d_y)$ for each design are reported in Table~\ref{tab:dimentions}.

\begin{table}[t]
\renewcommand{\arraystretch}{1.6}
    \begin{center}
    \caption{Dimensions of the one-dimensional finite linear array}
    \label{tab:dimentions}
    \footnotesize
    \begin{tabular}{|c|c|c|} \hline
        \multirow{2}{*}{Parameters} & Design-$1$ & Design-$2$ \\ \cline{2-3}
        & $\lambda/2$ spacing & $\lambda/4$ spacing \\ \hline
        Array dimension $(2a\times 2b)$ & $10.5\lambda\times0.5\lambda$ & $10.25\lambda\times0.5\lambda$ \\ \hline
        Square Patch dimension & $0.2385\lambda$ & $0.2192\lambda$ \\
        \hline      
        Probe position $(d_x,d_y)$ & $(0,0.0415\lambda)$  & $(0,0.0997\lambda)$ \\ \hline
       Total loaded elements $(N)$ &  $21$  & $82$ \\ \hline
       Substrate height ($h$) & \multicolumn{2}{c|}{$0.338$~mm } \\ \hline
    \end{tabular}
    \end{center}
\end{table}

\begin{figure}[t]
\centering
\includegraphics[width=3.45in,height = 4.1in]{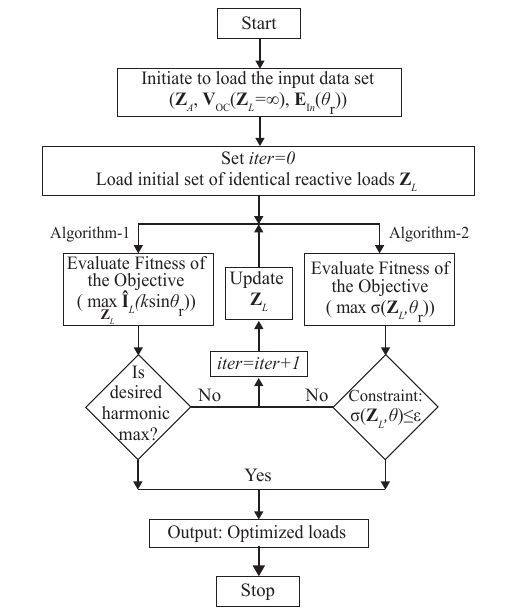}
\caption{Flow chart for the proposed algebraic optimization algorithms implemented in the MATLAB program for desired anomalous reflection.}
\label{fig:algorithms}
\end{figure}

\subsection{Objective Function: Desired Spatial Harmonic Maximization} \label{subsec:resultsobj2}

Based on the described main idea outlined in~Sec.~\ref{sec:methodology}, the natural optimization objective is to maximize the desired  harmonic within the spatial spectrum of the current distribution. This approach is aligned with the goal of perfect anomalous reflection, as this specific harmonic governs the reflection towards the receiver. The objective function of this approach is described as follows:
\begin{equation}
\label{Eq:obj_fun}
\mathcal{O}=\max_{\mathbf{Z}_L}
\{ |\widehat{\mathbf{I}}_L(k\sin\theta_{\rm {r}})|\}.
\end{equation}
Here, $\widehat{\mathbf{I}}_L(k_x)=\mathscr{F}\left\{\mathbf{I}_L (x)\right\}$ is the Fourier transform of load currents that maps the current distribution from the space domain to the tangential wavenumber or spatial frequency domain (Eq.~\eqref{Eq:transform}). By considering the harmonic at $k\sin\theta_{\rm {r}}$, we individually maximize the induced current at the desired wavenumber corresponding to the desired direction. This optimization is performed by continuously adjusting the load parameters. It is noteworthy to emphasize that, based on~Eq.~\eqref{Eq:obj_fun}, the objective function is independent of the incident angle and is solely dependent on the receiver's direction $\theta_{\rm r}$, a known parameter. In this current-based synthesis algorithm, the method begins by loading the pre-computed data set acquired from full-wave EM simulations. The objective function is maximized by varying the reactive loads $\mathbf{Z}_L$ of the array elements. Optimization is performed using initially assigned identical load reactances, with a randomly selected value.
The algorithm iterates by updating the reactive loads and evaluating the fitness of the objective function. The iterations continue until the desired harmonic reaches the maximum value, as depicted in Fig.~\ref{fig:algorithms}.

Further, the proposed method is tested on an example of two distinct structures with identical inclusions. The overall size of the structures is maintained identical in the $y$-direction, while the spacing is reduced by a factor of 2 for the second design. The first array comprises a single row with $21$ elements,  as depicted in~Figs.~\ref{fig1:conceptual_theory}(c)~\&~\ref{fig1:conceptual_theory}(d), where $p=1$ and $q=21$, with a fixed inter-element spacing at $d = \lambda/2$. The overall patch antenna array spans $10.5\lambda\times0.5\lambda$. The receiver is located along $70^\circ$ relative to the normal direction of the RIS plane. 

\begin{figure}[tb]
\centering
\includegraphics[width=0.5\textwidth]{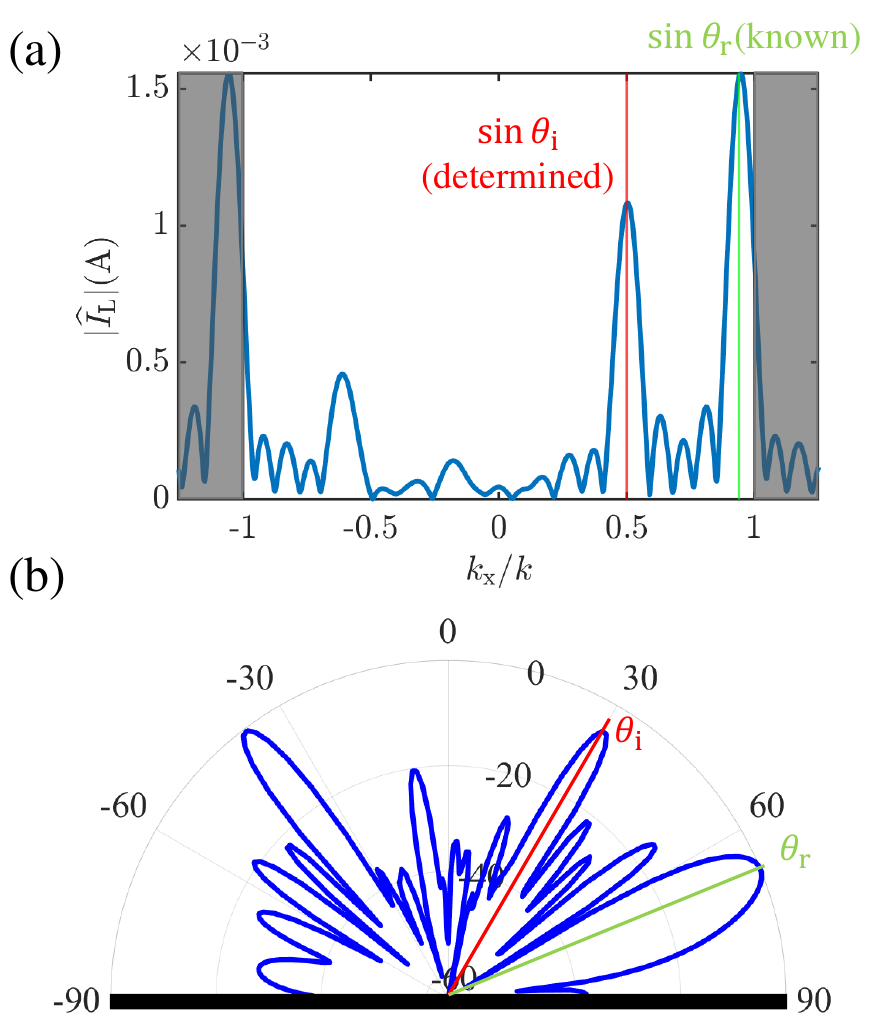}
\caption{(a) The amplitude distribution of the tangential wavenumbers  (spatial frequencies) of the currents at the loads for the linear array of the anomalous reflector with a $\lambda/2$ element spacing (design~$1$).   The green and red vertical lines correspond to the spatial frequency harmonics associated with the known desired reflection angle and the unknown incoming wave angle, respectively. Moreover, two vertical black lines depict the boundaries of the propagating harmonic range such that within the range, the harmonics are propagating because $|k_x|<k$, and outside the range (shown with the gray shade), the harmonics are evanescence modes. (b) Illustration of a   simulated SCS pattern $(\phi=0)$ in $xz$-plane for the different angles. The blue line demonstrates the normalized SCS for the optimized load values, reported in dBsm (decibels relative to a square meter).}
\label{fig:curr_opt}
\end{figure}

Figure~\ref{fig:curr_opt} depicts the results of the optimized RIS. As expected, two dominant peaks appear within the range of propagating waves ($|k_x/k|< 1$), as shown  in Fig.~\ref{fig:curr_opt}(a). The dominant peak aligns with the spatial frequency that is related to the desired reflection direction (i.e., $\theta_{\rm r}=70\degree$), while the secondary peak, with a lower amplitude, corresponds to the spatial frequency associated with the unknown incident angle. 
The estimated AoA shows accurate detection at an incidence angle of $\theta_{\rm i}=30\degree$ (which was chosen arbitrarily). Furthermore, the normalized SCS pattern is plotted in~Fig.~\ref{fig:curr_opt}(b) to further assess the performance.
Consequently, we achieved two key outcomes: first, we effectively rerouted the power toward the desired direction where the receiver is positioned. Secondly, by analyzing~Fig.~\ref{fig:curr_opt}(a), the tangential wavenumber associated with the lower peak allows us to determine the AoA. In other words, although the primary objective was to maximize the power reflected toward the receiver, the spatial frequency spectrum simultaneously enables accurate AoA detection and estimation. We note, however, that  the parasitic scattering in the $\lambda/2$-spacing  configuration is not effectively suppressed (the main parasitic direction is near $-40^\circ$ with normalized tangential wavenumber $k_x/k = -0.64$).

\begin{figure}[tb]
     \centering
    \includegraphics[width=0.5\textwidth]{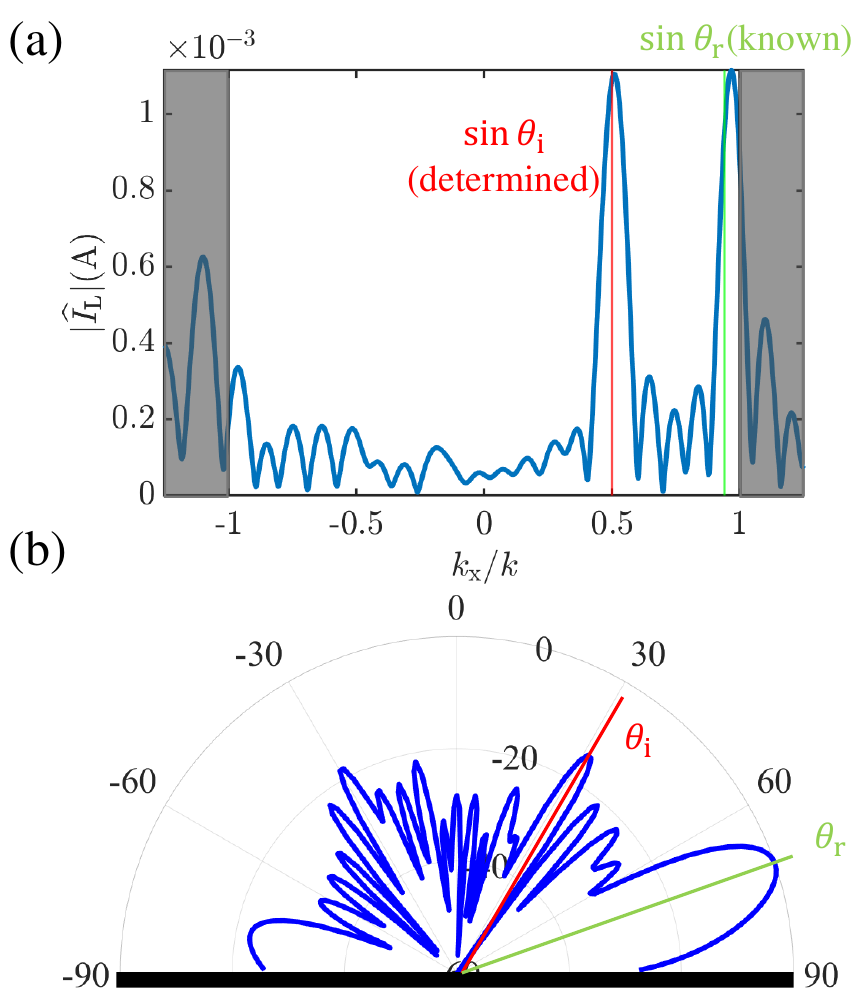}
    \caption{(a) The amplitude distribution of the spatial spectrum as a function of the tangential wavenumber, or spatial frequency, is shown for the linear array of the anomalous reflector with a $\lambda/4$ element spacing (design~$2$). All interpretations provided for~Fig.~\ref{fig:curr_opt}~(a) are valid here. (b) Illustration of 1-D CST simulated SCS pattern $(\phi=0)$ in $xz$-plane for various angles. The blue curve shows the pattern with the optimized load values. The normalized SCS values are reported in dBsm.}
    \label{fig:curr_opt_qw}
\end{figure}

For the second design, while all the other features remain identical, the element spacing is reduced to a quarter wavelength ($d = \lambda/4$). It comprises a dual row with $82$ elements, where $p=2$ and $q=41$, with an overall dimension of $10.25\lambda\times0.5\lambda$ with a fixed $\lambda/4$ inter-element spacing. The design-$2$ results are depicted in~Fig.~\ref{fig:curr_opt_qw}. Figure~\ref{fig:curr_opt_qw}(a) presents the amplitude of the Fourier transform of spatial current as a function of the spatial frequency (the normalized tangential wavenumber), and~Fig.~\ref{fig:curr_opt_qw}(b) shows the normalized SCS distribution. 

It is worth highlighting that based on the result on~Fig.~\ref{fig:curr_opt_qw}(a), the strongest peak is located near the unknown incident angle, $\theta_{\rm i}=30\degree$. This indicates that the optimized array shows  superdirective properties due to its subwavelength configuration. In other words, to fully eliminate the specular reflection, the obtained near-perfect solution redirects all the energy to the desired direction. Figure~\ref{fig:curr_opt_qw}(b) validates this prediction, as the amplitude level in the specular and undesired directions are significantly lower compared to the SCS pattern in Fig.~\ref{fig:curr_opt}(b). To further substantiate this claim, it is noteworthy to assess the efficiency by comparing these two structures and their respective solutions in terms of the optimized SCS patterns. For periodic structures, the efficiency of each harmonic was defined in the previous works as the ratio between the reflected harmonic and the amplitude of the incident wave. In contrast, the efficiency for the aperiodic structures was introduced differently~\cite{li2023tunable,Mostafa24SensingWires}. This  definition is based on the far-field patterns and can be presented as
\begin{equation}
\zeta = \frac{\vert \mathrm{\bf E}^{\rm far}(\theta_{\rm r}, \mathbf{Z}_L) \vert ^2} {\vert \mathrm{\bf E}_{\rm refer}^{\rm far} (\theta_{\rm r}) \vert ^2},
\label{eq:efficiency}
\end{equation}
where $ E^{\rm far}(\theta_{\rm r}, \mathbf{Z}_L)$ is the far-zone amplitude of the scattered field  in the direction of $\theta_{\rm r}$ for the array with the load impedances $\mathbf{Z}_L$. In the denominator, $\mathrm{\bf E}_{\rm refer}^{\rm far} (\theta_{\rm r})$ is the far-field amplitude in the reference case that  assumes a theoretically perfect, power-conserving anomalous reflector that supports the ideal current of an infinite anomalous reflector but over a finite area of the actual array. In this case, the scattered field from the structure can be found analytically~\cite{MacroscopicARM2021}~(for further details, refer to Appendix~\ref{sec:appendix_A}):
\begin{align}
    \mathrm{\bf E}_{\rm refer}^{\rm far} (\theta) &= \frac{j k e^{-jk|\mathbf{r}|}}{4 \pi |\mathbf{r}|} E_0 S\Bigg[\left( \cos(\theta) - \cos(\theta_{{\rm i}}) \right) \, {\rm sinc}(ka_{{\rm ef}}) \notag \\ 
    & \hspace{1cm} + r_n\left( \cos(\theta_{{\rm r}}) + \cos(\theta) \right) {\rm sinc}(ka_{{\rm ef}n})\Bigg],
\label{Eq:scatt_ref}
\end{align}
where $a_{\rm{ef}} = (\sin{\theta} - \sin{\theta_{\rm{i}}})a$ and $a_{{\rm ef}n} = (\sin{\theta} - \sin{\theta_{\rm{r}}})a$ characterize the effective sizes of the metasurface for the incident and reflected waves, respectively. Moreover, $\mathbf{r}$ is the position vector from the origin to the observation point, $S$ denotes the aperture area, and  $\rm sinc$ is the sinc function [${\rm sinc}(x)=\frac{\sin{x}}{x}$]. Value $r_n$ in~Eq.~\eqref{Eq:scatt_ref} is the macroscopic reflection coefficient that is defined as the ratio of the amplitude of the perfectly reflected wave in the desired direction to the amplitude of the incident wave. This ratio can be derived from the principle of power conservation, stating that the total power density incident on the infinite boundary (in the absence of active elements) must be equal to the power density reflected in the desired direction, yielding $r_n=\frac{\sqrt{\cos{\theta_{\rm{i}}}}}{\sqrt{\cos{\theta_{\rm{r}}}}}$ in our case. 

Based on this definition, the power balance for the reference case is perfect, resulting in the highest directivity for non-superdirectve solutions. The term "superdirective" indicates solutions when the computed efficiency using (\ref{eq:efficiency}) exceeds unity. Therefore, to have a comparison between the different designs, using~Eq.~\eqref{eq:efficiency}, the efficiency for a half-wavelength spacing ($d = \lambda/2$) is $89.73\%$, while for a quarter-wavelength spacing ($d = \lambda/4$), the efficiency rises to $177\%$, as reported in Table~\ref{tab:eff}. The observed efficiencies above 100\% are due to the finite aperture size and the quarter-wavelength spacing between elements, enabling sub-wavelength current distribution optimization in a densely loaded array.

\begin{table}[t]
\renewcommand{\arraystretch}{1.6}
    \begin{center}
    \caption{Optimized efficiency comparison for two objective functions with $\lambda/2$ and $\lambda/4$ element spacing}
    \label{tab:eff}
    \footnotesize
    \begin{tabular}{|c|c|c|} \hline
    \multirow{2}{*}{Maximizing objective function} & Design $1$ & Design $2$ \\ \cline{2-3}
    & $\lambda/2$ spacing Eff. & $\lambda/4$ spacing Eff.\\ \hline
    $|\widehat{\mathbf{I}}_L(k\sin\theta_{\rm {r}})|$ & 89.73\% & 177\% \\ \hline
    $\mathbf{\sigma}(\mathbf{Z}_L, \theta_{\rm r})$ & 96.65\% & 211\% \\ \hline
    \end{tabular}
    \end{center}
\end{table}

\begin{figure}[tb]
\centering
\includegraphics[width=0.5\textwidth]{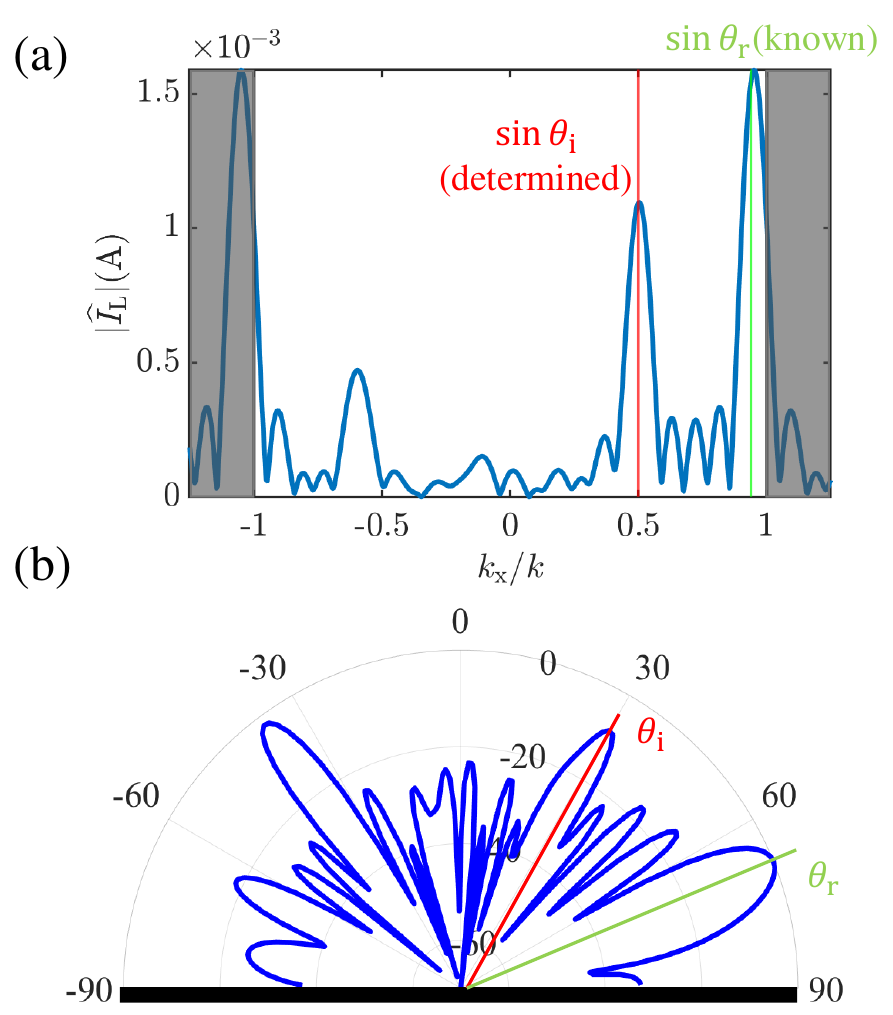}
\caption{The optimization results for maximizing the far-field radiation of the linear
array anomalous reflector with $\lambda/2$ element spacing (design 1) are presented. All interpretations provided for~Fig.~\ref{fig:curr_opt}~(a) remain valid here. (a) The spatial spectrum of the optimized current. (b)  Illustrates the far-zone   simulated   SCS pattern $(\phi=0)$ distribution after 
the optimization. The normalized SCS values are reported in dBsm.}
\label{fig:farfield_opt_hw}
\end{figure}

\begin{figure}[tb]
     \centering
    \includegraphics[width=0.5\textwidth]{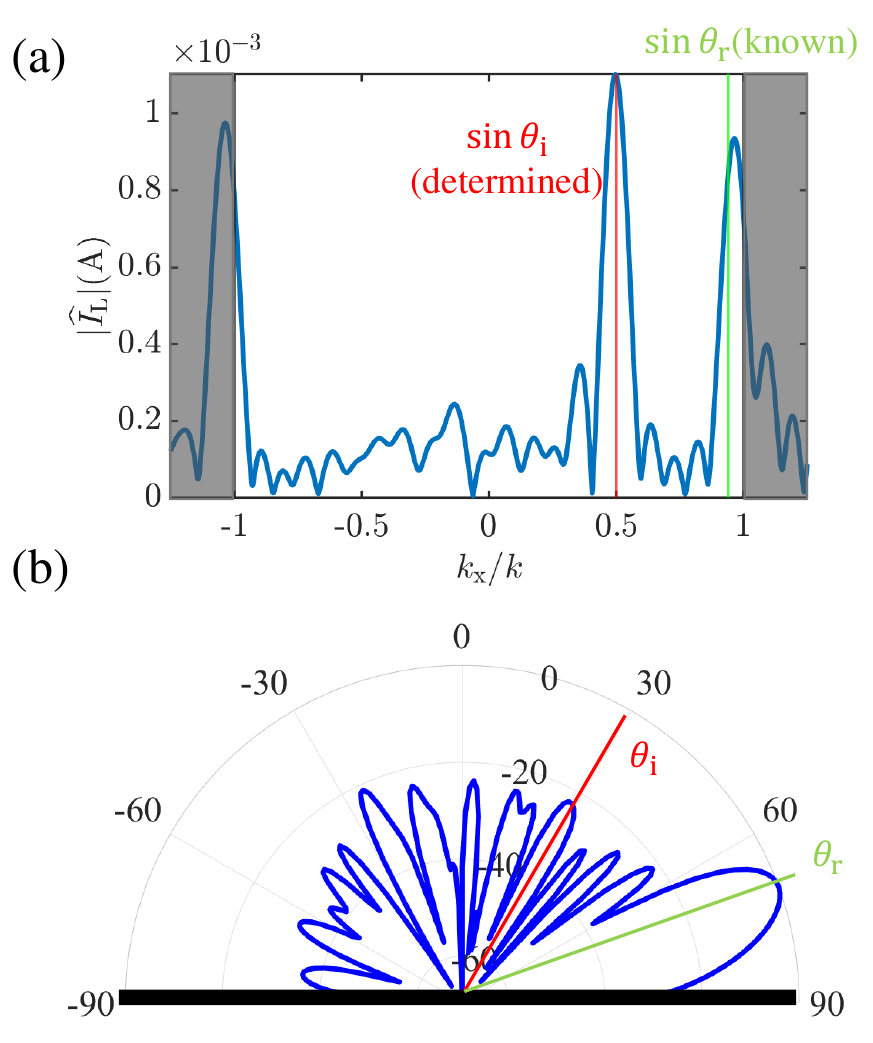}
    \caption{The optimization results for maximizing the far-field radiation of the anomalous reflector with $\lambda/4$ element spacing (design 2) are presented. All interpretations provided for~Fig.~\ref{fig:curr_opt}~(a) still apply. (a) The spatial spectrum of the optimized current. (b) Illustrates the far-zone simulated SCS pattern $(\phi=0)$ distribution of the post-optimization 
        loads. The normalized SCS values are reported in dBsm.}
    \label{fig:farfield_opt_qw}
\end{figure}

\subsection{Objective Function: Maximization of the Far-Zone Field in the Desired Direction}

Next, the objective function is changed to explore the potential for achieving more super-directive solutions. Instead of maximizing the current harmonic in the spatial spectrum responsible for the desired direction, the synthesis focuses on directly maximizing the reradiated power in the far zone in the direction of the receiver position. Based on this requirement, the following objective function will be employed in the optimization process:
\begin{equation}
\label{Eq:far_obj}
\mathcal{O} = \max_{\mathbf{Z}_L} \left \{ \mathbf{\sigma}(\mathbf{Z}_L, \theta_{\rm r}) \right \}.
\end{equation}
In the far-field-based synthesis algorithm, the technique focuses on maximizing the SCS in the desired anomalous direction. The optimization procedure similarly begins by loading the pre-computed dataset with initial phase-gradient reactive loads, iteratively synthesizing them into optimized reactive loads. In this method, the fitness function is based on maximizing the SCS in the desired direction, $\mathbf{\sigma}(\mathbf{Z}_L, \theta_{\rm r})$, while ensuring that it does not exceed a predefined threshold $\epsilon$ for specular and undesired scattering, where $\epsilon$ is a small positive constant. Imposing constraints on the synthesis technique is crucial to effectively suppress unwanted deflections in specular reflections and other unwanted directions. The algorithm iterates until the optimal solution of the criteria is met, as illustrated in Fig.~\ref{fig:algorithms}.

The optimization outcomes for $\lambda/2$ spacing are illustrated in~Fig.~\ref{fig:farfield_opt_hw}. As expected from the previous optimization algorithm, the spectrum in Fig.~\ref{fig:farfield_opt_hw}(a) for the half-wavelength element distance accurately reveals two peaks: one with higher amplitude in the desired direction and the second indicating the direction of the incident wave, which in this example is $30\degree$ incidence. Similar to the results with the current-based objective function, Fig.~\ref{fig:farfield_opt_hw}(b) shows that most of the power is deflected toward the desired direction. The radiation pattern of the optimized case  exhibits a lower amplitude in the specular and undesired directions. However, a small SCS amplitude remains in that direction even after optimization. Nonetheless, the linear array structure successfully redirects the power to the desired direction while enabling AoA detection. 

However, as previously mentioned, to achieve an effective solution, the most straightforward approach is to reduce the element spacing. Therefore, similar to the earlier study in Section~\ref{subsec:resultsobj2}, design~$2$ with a quarter wavelength element spacing is developed and re-optimized. A superdirective solution was again identified with $\lambda/4$ distancing, as depicted in Fig.~\ref{fig:farfield_opt_qw}. Figure~\ref{fig:farfield_opt_qw} (a) displays two peaks within the propagation borders, marked by vertical black lines. The peak order differs from the normal cases, like $\lambda/2$ spacing designs. This should be elaborated on in detail. Since the solution offers a superdirective structure, the highest amplitude for the specular direction (near the red vertical line) results from the complete suppression of direct reflection from the ground plane. 

To further confirm this claim, the efficiency parameter is calculated as defined in Eq.~\eqref{eq:efficiency}. The efficiency for the $\lambda/2$ spacing is $96.65\%$, while for the $\lambda/4$ spacing, the efficiency reaches $211\%$. These efficiency values are summarized in Table~\ref{tab:eff} for reference.

\section{Conclusion}\label{sec:conclusion}

A numerically efficient dual-functional RIS for AoA sensing and anomalous reflection utilizing realistic patch antenna arrays has been presented. This approach can be broadly generalized to accommodate arbitrary structural geometries and incident wave polarizations. Moreover, it remains applicable in scenarios involving multiple incident wave excitations. The main idea is to optimize the Fourier transform of the measured spatial current or voltage distribution at the antenna array terminals for a specific spatial harmonic responsible for the reflection toward the desired receiver angle while simultaneously detecting the incident angle. The numerical validation outcomes effectively control the extreme deflection angles for desired anomalous reflection using circuit-based arithmetical optimization of receiving and scattering array antenna theory, which is efficient compared to traditional full-wave EM simulation-based techniques.

The proposed AoA concept eliminates the necessity for additional circuitry RF chains and active sensing antennas, compared with existing alternative AoA detection solutions, and offers simpler hardware as a passive reflector reconfigured electronically offers dynamic control and improves AoA detection accuracy. This study validated the proposed concept through practical microstrip patch arrays for three-dimensional models at mmWave frequencies under TE polarization. The core theoretical concept is general, allowing the approach to be applied across diverse frequencies and configurations, ensuring broad design applicability. In future research, we plan experimental verification of the proposed approach.

\appendices
\section{Definition of reference case in efficiency calculation}\label{sec:appendix_A}

\begin{figure}[t]
\centering
\includegraphics[width=2.6in,height = 0.8in]{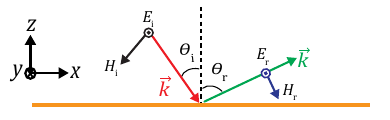}
\caption{The side view of the schematic structure illustrates the illumination of the TE-polarized plane wave, and the incidence plane is the $xz$-plane.}
\label{fig:schematic_eff}
\end{figure}

This section will summarize the derivation of the reference case regarding the efficiency calculation when a periodic finite structure is assumed as the reference. The assumed boundary for the case is shown in Fig.~\ref{fig:schematic_eff} where the incoming electric wave has TE polarization and the vector can be presented with $\mathbf{E}_{\rm{i}}(x, z) = E_0 \, e^{-jk(\sin{\theta_i} x - \cos{\theta_i} z)} \hat{a}_{y}$. As we mentioned earlier, the perfect reflector should be able to reflect all the incoming power in the desired direction. It means that the relation between the amplitude of the reflected wave ($E_{\rm{r}}$) with respect to the amplitude of the incoming wave ($E_{0}$) is
\begin{equation}
\label{Eq:refl_elec_amp}
E_{\rm{r}} = E_{0} \frac{\sqrt{\cos{\theta_{\rm{i}}}}}{\sqrt{\cos{\theta_{\rm{r}}}}}.
\end{equation} 
Therefore the reflected electric field can be constructed $\mathbf{E}_{\rm{r}}(x, z) = E_r \, e^{-j(k \sin{\theta_i} x + k \cos{\theta_i} z)} \hat{a}_{\rm{y}}$. In addition, the magnetic field can be simply obtained as $\mathbf{H}_{\rm{i,r}} = \frac{1}{\eta} \hat{k}_{\rm{i,r}} \times \mathbf{E}_{\rm{i,r}}$ where $\hat{k}_{\rm{i,r}}$ is the direction of propagation. Next, the magnetic and electric current densities can be achieved using \cite{asadchy2017flat, MacroscopicARM2021}
\begin{align}
    \mathbf{J} &= \hat{{z}} \times \left( \mathbf{H}_{\rm{i}} + \mathbf{H}_{\rm{r}} \right) \notag \\
               &= \frac{1}{\eta} \left( E_0 \cos{\theta_i} e^{-j(k \sin{\theta_i} x)} \right. \notag \\
               &\hspace{1cm} \left. + E_r \cos{\theta_r} e^{-j(k \sin{\theta_r} x)} \right) \hat{{\rm{y}}}
    \label{eq:electric_current_density}
\end{align}
\begin{align}
    \mathbf{M} &= -\hat{{z}} \times \left( \mathbf{E}_{\rm{i}} + \mathbf{E}_{\rm{r}} \right) \notag \\
               &= \left( E_0 e^{-j(k \sin{\theta_i} x)} + E_r e^{-j(k \sin{\theta_r} x)} \right) \hat{{\rm{x}}}
    \label{eq:magnetic_current_density}
\end{align}
It is important to note that only the tangential components of the electric and magnetic fields, both the incident and reflected, contribute to the formation of currents on the surface. Using these currents, the scattering characteristics of the aperture can be obtained using~Eq.~\eqref{eq:Fourier_aperture}. In other words,~Eq.~\eqref{eq:Fourier_aperture} is the surface integral that can be interpreted as the spatial Fourier transform of the surface current distribution \cite{asadchy2017flat, MacroscopicARM2021}.
\begin{equation}
    \mathbf{E}_{\rm{sc}}(\mathbf{r}) = \frac{j k e^{-jk|\mathbf{r}|}}{4 \pi |\mathbf{r}|} \left[ \hat{\mathbf{r}} \times \int_{\rm{S}} \left( \mathbf{M} + \eta \hat{\mathbf{r}} \times \mathbf{J} \right) e^{j k \mathbf{r} \cdot \mathbf{r}'} dx'dy' \right]
    \label{eq:Fourier_aperture}
\end{equation}
Replacing~Eqs.~\eqref{eq:electric_current_density}, ~\eqref{eq:magnetic_current_density} to~Eq.~\eqref{eq:Fourier_aperture} and applying some simplification the final representation for the far-zone scattered field will be~Eq.~\eqref{Eq:scatt_ref}.

\ifCLASSOPTIONcaptionsoff
  \newpage
\fi

\bibliographystyle{IEEEtran}
\bibliography{IEEEabrv,Bibliography}

\end{document}